\documentclass[twocolumn,preprintnumbers,amsmath,amssymb]{revtex4}
\usepackage{graphicx}% Include figure files
\usepackage{dcolumn}% Align table columns on decimal point
\usepackage{bm}% bold math

\newcommand\beq{\begin{eqnarray}}
\newcommand\eeq{\end{eqnarray}}

\newcommand\bal{\begin{align}}
\newcommand\eal{\end{align} }

\newcommand{\mybar}[1]%
        {\kern 0.6pt\overline{\kern -0.6pt#1\kern -0.6pt}\kern 0.6pt}
\begin{document}

\preprint{INT-PUB-09-029}

%\title{Reduction of Baryon Noise Through Partial Quenching}
\title{Reducing Baryon Noise in Lattice QCD\\Through Partial Quenching}% Force line breaks with \\

\author{Amy N. Nicholson}%
 \email{amynn@u.washington.edu}
\affiliation{%
Institute for Nuclear Theory, Box 351550, Seattle, WA, 98195-1550}%


\begin{abstract}
The study of nuclear physics using lattice QCD is hindered by an exponentially large signal-to-noise problem which is conventionally alleviated by raising the quark masses to unphysically high values. We propose a novel form of partial quenching for calculations involving nucleons in which the sea quark masses are taken to be smaller than the valence quark masses. It is shown that lowering the sea quark masses toward their physical values actually improves signal-to-noise. An optimized approach to the physical point in the ($m_{s}$, $m_{v}$) plane is proposed, with a full analysis of the cost benefit. Improvements in computing time of $\sim 10^{2(A-1)}$, where A is the number of nucleons in the system, are shown to be possible.
\end{abstract}

\maketitle

\section{\label{sec:level1}Introduction}

Lattice QCD provides a promising tool for the calculation of properties of nuclear systems from first principles. In particular, one goal is to use lattice QCD to gain access to quantities for which we have little or no experimental data, such as the three neutron interaction, necessary as input for many nuclear models, or hyperon-baryon interactions, which may have relevance to the equation of state for neutron stars. As improvements in computing power and algorithms continue to allow more precision in lattice calculations, we are entering an exciting era in which the calculation of properties of multiple baryon systems is becoming possible, as evidenced by the recent appearance of the first study of three baryons \cite{Beane:2009gs}. However, calculations involving baryons with light valence quarks still suffer from an exponential degradation in time of signal-to-noise, resulting in large errors. To overcome this problem will require enormous computational resources as the number of baryons is increased. Creative methods for reducing the signal-to-noise ratio (SNR) are necessary if we wish to further explore nuclear physics on the lattice (see, e.g., \cite{Bedaque:2007pe}). In this paper, we investigate the quark mass dependence of the SNR and present a new approach which will greatly reduce the computational time associated with these calculations.

A shared characteristic of most lattice calculations to date is the use of unphysically large quark masses. Typically this is done because as one lowers the quark masses, critical slowing down of the algorithms employed to invert the Dirac matrix occurs. For calculations involving nucleons, the SNR is also improved at larger quark masses. Based on the elegant argument by Lepage \cite{Lepage:1989hd} - outlined in Sec. II - one can show that the SNR for a correlator of an operator consisting of interpolating fields for A nucleons is approximately $\sqrt{\mathcal{N}}e^{-A (M_N -\frac{3}{2} m_{\pi})\tau}$, where $M_N$ is the mass of the nucleon, $\tau$ is the Euclidean time separation of source and sink, and $\mathcal{N}$ is the number of measurements made. For larger quark masses, explicit chiral symmetry breaking is more severe, and the difference $M_N -\frac{3}{2} m_{\pi}$ is smaller, thus improving the SNR. 

Raising the quark masses results in systematic errors, and extrapolation must be performed to obtain physical answers. In an attempt to simultaneously avoid critical slowing down and reduce systematic errors, partial quenching, in which the valence quark masses are taken to be smaller than the sea quark masses, has been employed in the mesonic sector. However, it has not been clear whether this method would be beneficial in the baryonic sector due to a reduced SNR.

A more careful study of the quark mass dependence reveals that an unconventional form of partial quenching, in which the valence quark masses are taken to be \emph{larger} than the sea quark masses, actually improves the SNR. In addition, recent results indicate that it is propagator production and contractions which consume the largest amount of computing time for baryon calculations \cite{Beane:2009ky}, contrary to the mesonic sector. In this paper we investigate the quark mass dependence of an array of factors affecting the precision of lattice calculations involving nucleons, and propose that the ideal program for approaching the physical limit in baryonic calculations is to calculate at physical sea quark mass and extrapolate in the valence quark mass only.

\section{\label{sec:level1}Signal to Noise Estimates}

Conventionally, hadron properties are computed on the lattice by considering correlators of the form $G(\tau) = \langle 0 | \mathcal{B}(\tau)\mathcal{B}^\dagger (0) |0 \rangle $, where $\mathcal{B}$ has some overlap with the state of interest. After analytically integrating out the fermions, one is left with a new operator, $\mathcal{O}(\tau;A)$, consisting of quark propagators from $t=0$ to $t=\tau$, which is a function of the gauge fields. For large time separation, the correlator of this object will project out the lightest state produced by the operator, $G \propto e^{-E_{0} \tau} $. From this, one can extract the energy of the state ($E_{0}$).

Here, we outline the arguments presented by Lepage to estimate the SNR for correlators calculated on lattices with anti-periodic temporal boundary conditions and an infinite time extent. Since the correlators are approximated by sampling $\mathcal{N}$ independent gauge configurations,
\beq
G_{\mathcal{N}}(\tau) = \frac{1}{\mathcal{N}} \sum_{i=1}^{\mathcal{N}} \mathcal{O}(\tau;A_i)
\eeq
the SNR ($\mathcal{R}$) can be computed for large $\mathcal{N}$ using the central limit theorem,
\beq
\mathcal{R} &\approx& \sqrt{\mathcal{N}} \frac{G_{\mathcal{N}}}{\sigma} , \cr
\sigma^2 &\equiv& \langle | \mathcal{O} (\tau;A)|^2\rangle - |\langle \mathcal{O} (\tau;A)\rangle |^2 .
\eeq
At large times the correlator in the first term of $\sigma^2$ will project out the lightest state produced by $\mathcal{O} \mathcal{O}^\dagger$, so phenomenological knowledge about the strong interactions can be applied to make predictions about the SNR. In particular, if $\mathcal{B}$ is an operator for producing A nucleons, then $\mathcal{O}\mathcal{O}^\dagger$ will consist of 3 A quark and 3 A antiquark propagators from $t=0$ to $t=\tau$. The lightest state projected out by this operator will be 3 A pions at rest, so we would estimate that our SNR is given by
\beq
\mathcal{R} \approx \sqrt{\mathcal{N}} e^{-A (M_N - \frac{3}{2} \tilde{m}_{\pi}) \tau} .
\eeq
Here, $\tilde{m}_{\pi} \approx m_{\pi}$ is the mass of the pion in the partially quenched theory where valence quark annihilation is disallowed. 

Since for physical masses, $M_N-\frac{3}{2} m_{\pi}\sim 730$ MeV, a very large number of measurements is required for large $\tau$ ($\tau \gtrsim 1$ fm) in order to see a statistically significant signal. At shorter time separations, the correlator will be contaminated by excited states. In practice, one must use a finite time extent, and Eq. [3] has recently been shown to give a good approximate upper bound on the SNR for $\tau \lesssim 2$ fm on a lattice with a $4.5$ fm time extent \cite{Beane:2009ky}. Above this, backward propagating states must be taken into account, and the SNR is expected to be much worse. We will concentrate on the range for which the Lepage expression holds (Eq. [3]), since it is here that measurements are most likely to be made.

\section{SNR in the ($m_s$, $m_v$) plane}
The main source of the signal-to-noise problem is spontaneous chiral symmetry breaking, which causes the pions to be light. Thus, it is expected in general that the SNR will improve with heavier quark masses. More specifically, it is the ability of $\mathcal{O}\mathcal{O}^\dagger$ to produce light pions which affects the SNR. So one might ask whether it is necessary to raise all of the quark masses in order to improve the SNR, or only the valence quarks associated with the interpolating field. The key to quantifying the effect on the SNR of changing sea and quark masses independently lies in partially quenched chiral perturbation theory (PQ$\chi$PT). 

The mass of the nucleon in SU(2) PQ$\chi$PT has been calculated to order $\mathcal{O}(m_q^{2})$ in \cite{Tiburzi:2005na}. We have,
\begin{widetext}
\beq
M_N &=& M_{0}+\alpha m_{vv}^2+\beta m_{ss}^2 -\frac{1}{16\pi f^2} \lbrace \frac{g_A^2}{12}[-7m_{vv}^3+16m_{vs}^3+9m_{vv}m_{ss}^2] + \frac{g_1^2}{12}[-19m_{vv}^3+10m_{vs}^3+9m_{vv}m_{ss}^2] \cr
& &+ \frac{g_1 g_A}{3}[-13m_{vv}^3+4m_{vs}^3+9m_{vv}m_{ss}^2]\rbrace -\frac{4g_{\Delta N}^2}{3}\frac{\mathcal{F}(m_{vv},\Delta,\mu)}{(4\pi f)^2}-\frac{4g_{\Delta N}^2}{3}\frac{\mathcal{F}(m_{vs},\Delta,\mu)}{(4\pi f)^2} + \mathcal{M}_4,
\eeq
where
\beq
\mathcal{F}(m,\delta,\mu) = (m^2-\delta^2)\left[\sqrt{\delta^2-m^2}\log \left(\frac{\delta-\sqrt{\delta^2-m^2+i\epsilon}}{\delta+\sqrt{\delta^2-m^2+i\epsilon}}\right)-\delta \log \left(\frac{m^2}{\mu^2} \right) \right] 
-\frac{1}{2}\delta m^2 \log \left(\frac{m^2}{\mu^2}\right)-\delta^3 \log \left(\frac{4\delta^2}{\mu^2}\right),
\eeq
and $\mathcal{M}_4$ contains the $\mathcal{O}(m_q^2)$ terms, presented in \cite{Tiburzi:2005na}.
For the pion we have \cite{Sharpe:2000bc},
\beq
M_\pi^2 = m_{vv}^2 \left( 1+\frac{16}{f^2}(2 L_6-L_4)m_{ss}^2 +\frac{8}{f^2}(2L_8-L_5)m_{vv}^2+\frac{1}{32 \pi^2f^2}\left[ m_{vv}^2-m_{ss}^2+(2m_{vv}^2-m_{ss}^2)\log \frac{m_{vv}^2}{\Lambda_{\chi}^2}\right]\right)
\eeq
\end{widetext}
Here, $m_{qq^{'}}^2 = B_0 (m_q + m_{q^{'}})$, where $B_0 \sim 2.4$ GeV, and for physical pions, $m_q \approx m_{q^{'}}\sim 4$ MeV. The parameters in the expression for the pion mass are given in \cite{Allton:2008pn}. Fits of the parameters for the nucleon mass up to $\mathcal{O}(m_q^{3/2})$ were performed to the results presented in \cite{Ohki:2008ff} for values of the pion mass below 400 MeV. The unknown parameters contained in the $\mathcal{O}(m_q^2)$ terms for the nucleon mass expression were sampled from gaussian distributions and used as a measure of the systematic uncertainty. Note that, due to the $m_{vs}^3$ and $m_{vv}m_{ss}^2$ terms, changing the sea quark masses affects the nucleon mass at $\mathcal{O}\left(m_q\right)$, while the pion mass is unchanged until $\mathcal{O}\left(m_q^2\right)$.

Inserting these expressions into Eq. (3), and normalizing with respect to the SNR for physical quark masses, gives the result shown in Fig. 1 for systems with up to 4 nucleons. 
\begin{figure}
\includegraphics[width=\linewidth]{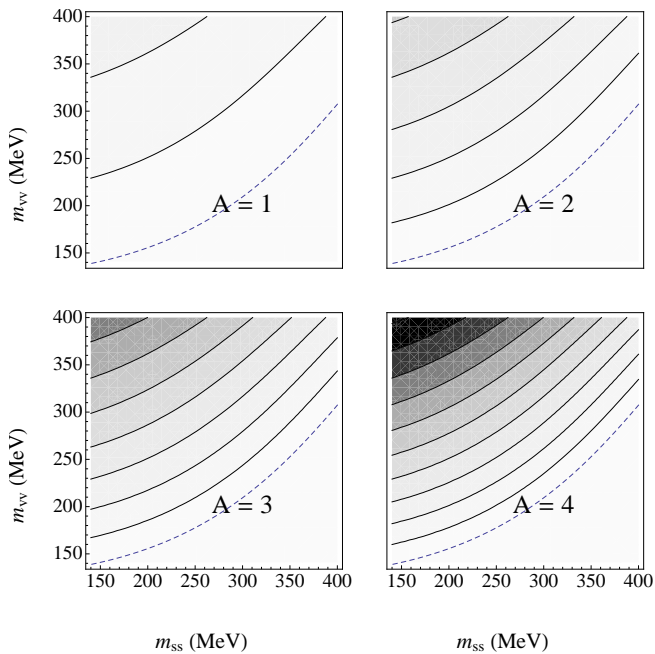}
\caption{\label{fig:epsart}(Color online) Contour plot in the $(m_{ss},m_{vv})$ plane of the following ratio of SNRs at $\tau = 1.5$ fm for $A=1$ to 4: SNR$(m_{ss},m_{vv})/$SNR$(m_{ss}=m_{vv}=140$ MeV). Each contour represents a factor of two with respect to the dashed curve, and darker contours correspond to larger ratios.}
\end{figure}
We see that the best value for the SNR occurs at high valence quark mass and low (physical) sea quark mass, with improvements over heavy sea quarks of about 3 times for a single nucleon, and 100 times for 4 nucleons. The reason for these improvements can be seen by inspecting the mass difference governing the SNR in Eq. (3). Keeping the sea quark mass at its physical value and raising $m_{vv}$ to $400$ MeV only raises the nucleon mass by about $130$ MeV, while the pion mass is raised by $240$ MeV, so that $M_N - \frac{3}{2} m_{\pi}$ has been reduced from $730$ MeV in the physical case to $500$ MeV.

Another expression to consider is the time at which exponential degradation of signal-to-noise sets in. It was shown in \cite{Beane:2009gs} that the standard projection of the nucleon state onto zero momentum introduces a volume suppression to the noise. For small time slices, this suppression can dominate, in which case the SNR no longer depends exponentially on the number of nucleons. The largest time for which this occurs was calculated to be,
\beq
t_{noise} \sim \frac{2}{2 M_N -3 m_{\pi}}\ln \left[\frac{m_{\pi}^3 L_x^3}{A^2}\right],
\eeq
where the spatial volume should be set by the Compton wavelength of the lightest pion in the system, $L_x \propto \frac{1}{m_{ss}}$. For $A=1$ we find, using the expressions above for the nucleon and pion masses at $m_{vv} = 400$ MeV, an approximate 50$\%$ increase in $t_{noise}$ when $m_{ss}$ is lowered from 400 MeV to 140 MeV.
\section{Cost of Calculation}
Based on Fig. 1, in order to optimize the SNR at a given valence quark mass, one should lower the sea quark masses to their physical values. This suggests that an ideal approach to the physical point would be to use physical sea quark masses for all measurements, and extrapolate only in the valence quark mass. This approach has the added benefits of requiring a single set of gauge field configurations to be produced for all measurements, as well as reduced systematic uncertainties.

To determine whether this approach will be beneficial in practice, one must include the cost of gauge field and propagator production. There are many factors involving the quark masses which affect computation time, including the cost of inverting the Dirac matrix, volume requirements, and number of independent sources permitted per gauge configuration.

The noise, including volume effects, can be approximated by \cite{Beane:2009gs}
\beq
\sigma^2 = \frac{1}{\mathcal{N}_g \mathcal{N}_{src}} \sum_{j=0}^A & &\left[\frac{(A!/j!)^2}{(m_{\pi} L_x)^{3(A-j)}}Z_{A-j} \right. \cr
& & \left. \times e^{-(2j M_N + 3(A-j) m_{\pi})\tau}\right] ,
\eeq
where $L_t$ is the temporal extent, $Z_i$ is the overlap onto the ith state, and $\mathcal{N}_{src} = \mathcal{N}_0 (A M_N)^4 L_x^3 L_t$ is the number of independent measurements which can be made on a single configuration. Based on the results in \cite{Beane:2009ky}, we have chosen the normalization $\mathcal{N}_0$ such that at $m_{ss} = m_{vv} = 390$ MeV we have 200 sources per configuration. To determine the cost of achieving a fixed SNR at a given quark mass, we found the number of gauge configurations necessary, $\mathcal{N}_g$, then multiplied by the cost of producing a single gauge configuration plus the measurements made on that configuration. To explore the effects of different cost functions we chose both domain wall and staggered fermion actions, and domain wall fermion propagators. The cost functions used are given in the Appendix.

Figure 2 shows the ratios of the total costs to perform calculations for $A = 1$ to 4 nucleons for a fixed 10$\%$ statistical error at $m_{vv} = 400$ MeV for two different values of sea quark mass: $m_{ss} = 400$ MeV versus $m_{ss} = 140$ MeV (circles). We also compared two separate approaches for extrapolation to the physical point to determine whether our proposal will be beneficial as one lowers the valence quark masses. First, we calculated the cost of producing gauge field configurations and propagators for equal sea and valence quark masses at 8 different values between 400 and 250 MeV, again at a fixed $10 \%$ statistical error. Then, we compared this with the cost of producing a single set of gauge configurations at physical sea quark mass, as well as propagators on these configurations at the same 8 values of valence quark masses between 400 and 250 MeV (squares). 
\begin{figure}
\includegraphics[width=\linewidth]{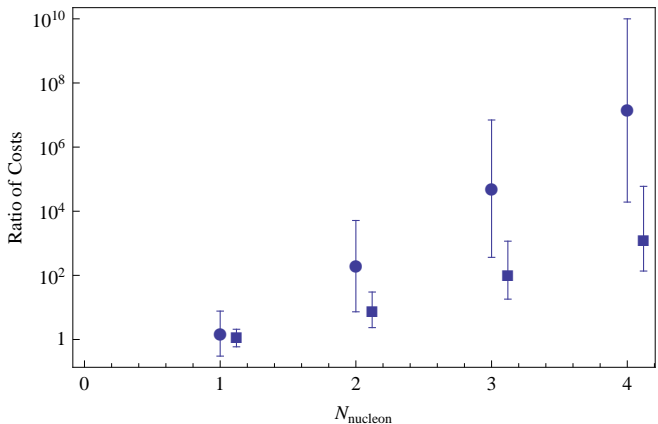}
\caption{\label{fig:epsart}(Color online) Log plot of the following ratio of costs at fixed SNR for calculations at different values of quark masses with domain wall fermions: Cost($m_{ss}= m_{vv} = 400$ MeV)$/$Cost($m_{ss} = 140$ MeV, $m_{vv} = 400$ MeV), as a function of the number of nucleons (circles). Also plotted is the ratio of costs for eight measurements using two different approaches to the physical point: Cost($250$ MeV $\leq m_{ss} = m_{vv}\leq  400$ MeV)$/$Cost($m_{ss} = 140$ MeV, $250$ MeV $\leq m_{vv} \leq 400$ MeV) (squares).}
\end{figure}

For this plot we chose $a = 0.093$ fm for the lattice spacing and domain wall fermion cost functions for both sea and valence quarks. For $A >1$, we found negligible sensitivity to changing either the lattice spacing or the cost function. Some sensitivity to the cost function was found for $A=1$. All error bars reflect the uncertainties in the HB$\chi$PT expressions for the masses, except for $A=1$ where the uncertainty due to the choice of cost function is also added in quadrature.

\section{Discussion}

For all but the single baryon case, it is clearly beneficial to calculate nucleon properties at physical sea quark masses. From the linear dependence of the log plots in Fig. 2 we see that improvements of $\sim 10^{2 (A -1)}$ for a single calculation and $\sim 10^{A-1}$ for a full approach to the physical point can be expected. It is still unclear whether our method would be beneficial for single nucleon calculations unless the gauge field configurations were already available. 

Note that we are only considering here a fixed statistical error. Lowering the sea quark masses will also help reduce the systematic errors associated with unphysical quark masses. This will be particularly beneficial for calculations in which theoretical tools for extrapolation to the physical point are less well developed. In addition, extrapolating to the physical point in only one quark mass will greatly reduce the proliferation of fit parameters usually associated with partial quenching. However, there may still be extra non-analytic terms introduced, which in principle will require more measurements to accurately determine their coefficients. 

A further consideration not addressed in this work is the possibility that lowering the sea quark mass will decrease the gap between the ground state and the first excited state. This could force one to make measurements at larger time separations in order to extract the ground state signal. This issue is highly dependent on the system one wants to consider, and can be improved by optimizing the interpolating field used, as well as the fitting technique. However, as observed in \cite{Beane:2009gs}, eliminating excited state effects from calculations involving multiple baryons can be difficult even for large sea quark masses. Because present day techniques may not be sufficient to extract many nuclear observables of interest, further study in these areas, particularly as calculations continue to move closer to the physical point, will be necessary.

\section{Conclusions}

We have shown that to optimize the SNR for nucleons in terms of valence and sea quark masses, one should choose heavy valence quark masses and physical sea quark masses. This choice not only improves the SNR as compared to the standard choice of heavy valence and sea quarks, but should also produce results which are closer to the physical case of interest, thus reducing systematic errors. These improvements become more significant as the number of baryons is increased, possibly making previously intractable calculations realistic in the near future.

\begin{acknowledgments}
The author wishes to thank D.B. Kaplan for initially suggesting this line of investigation, and for many discussions. The author also wishes to thank M. J. Savage and A. Walker-Loud for several useful discussions, and K. Orginos for helpful comments.
\end{acknowledgments}

\bibliography{PQQCDv7}% Produces the bibliography via BibTeX.

\appendix*

\section{Cost Functions}
Example cost functions for domain wall fermion propagators, domain wall fermion gauge field configurations, and staggered fermion gauge field configurations were taken from \cite{Bratt:2007vs}, \cite{Edwards:2007vs}, and \cite{Jansen:2008vs}, respectively. The parameters used in this work are given in Table 1.

\vspace{5mm}
\noindent Domain wall propagators:
\beq
\mbox{Cost} &\propto& \left( A+\frac{B}{m_{vv}} \right) L_x^3 L_t L_5
\eeq
Domain wall gauge configurations:
\beq
\mbox{Cost} &\propto& \left( \frac{L_x}{\mbox{ fm}} \right)^4 \left(\frac{L_t}{\mbox{ fm}}\right) L_5 \left(\frac{\mbox{MeV}}{m_{ss}}\right) \left(\frac{\mbox{ fm}}{a}\right)^7 \left(\frac{\mbox{MeV}}{m_{K}}\right)^2 \cr
& & \times \left( C_1 + C_2 \left(\frac{a}{\mbox{ fm}}\right)^3 \left(\frac{m_K}{m_{ss}}\right)^2 \right)
\eeq
Staggered gauge configurations:
\beq
\mbox{Cost} &\propto&  \left(\frac{20 \mbox{ MeV}}{m_s}\right)^{c_m}\left(\frac{L_x}{3 \mbox{ fm}}\right)^{c_L}\left(\frac{0.1\mbox{ fm}}{a}\right)^{c_a}
\eeq
Here, $m_s$ is the sea quark mass, and $m_K$ is the mass of a kaon containing a light sea quark. 
\begin{table}

\caption{\label{tab:table1} List of parameters for the given cost functions. }
\begin{tabular}{| c | c |}
\hline
&\\
A &  $7.98 \times 10^{-7}$ \hspace{.8 cm}\\
B &  $1.01 \times 10^{-6}$\hspace{.8 cm}\\
\hspace{.5 cm}$C_1$\hspace{.5 cm} & 0.01021\\
$C_2$ & 0.3226\\
$c_m$ &1 \\
$c_L$ &4 \\
$c_a$ &\hspace{1.2 cm}4\hspace{1.2 cm} \\
&\\
\hline
\end{tabular}
\end{table}

\end{document}